# Diffusion and heat conductivity in the weakly ionized plasma with power-law *q*-distributions in nonextensive statistics

Ji Xiaoxue and Du Jiulin

*Department of Physics, School of Science, Tianjin University, Tianjin 300072, China*

**Abstract -** We study the diffusion and thermal conductivity of charged particles in the weakly ionized plasma with the power-law *q*-distributions in nonextensive statistics and without the magnetic field. Electrons and ions have different *q*-parameters and temperature. We derive new expressions of the diffusion, thermal diffusion, thermal conductivity and thermoelectric coefficients of electrons and ions respectively in the plasma. It is shown that these transport coefficients depend significantly on the *q*-parameters in the power-law *q*-distributed plasma and thus they have different properties from those derived in the traditional statistics with a Maxwellian distribution.

## 1. Introduction

The classical Boltzmann-Gibbs (B-G) statistics can describe the statistical properties of the systems with short-range interactions. The systems with short-range interactions are generally extensive, and therefore B-G statistics is actually an extensive statistics and only has the exponential probability distributions. One of the remarkable examples for the extensive property is of the B-G entropy $S_{B\text{-}G}$. Namely, if $S(A)$ and $S(B)$ are the B-G entropies of two independent subsystems $A$ and $B$ respectively, the total B-G entropy of the system is $S_{B\text{-}G} = S(A)+S(B)$. However, the systems with long-range interactions, such as self-gravitating systems and plasmas, are generally nonequilibrium and complex states. They are not always extensive and in certain physical situations have the power-law distributions (non-exponential distributions), so the classical B-G statistics is not always suitable for describing the statistical properties of such systems.

In 1988, Tsallis proposed a nonextensive entropy defined [1] by

$$S_q = \frac{k_B}{1-q}(\sum_i p_i^q - 1) \tag{1}$$

where $k_B$ is Boltzmann constant, $p_i$ is the probability that the system under consideration is in its *i*th configuration, and $q$ a positive nonextensive parameter whose deviation from one describes the degree of nonextensivity of the system. The B-G entropy $S_B$ is obtained only when we take the limit $q\to 1$:

$$S_{B-G} = \lim_{q\to 1} S_q = -k_B \sum_i p_i \ln p_i. \tag{2}$$

In the expression (1), Tsallis entropy (or the q-entropy) is nonextensive and satisfies

$$S_q(A+B) = S_q(A) + S_q(B) + \frac{(1-q)}{k_B} S_q(A) S_q(B), \tag{3}$$

and the extensivity is recovered as the limit $q\to 1$.

Based on Tsallis entropy, the classical B-G statistics can be generalized to a new statistics, nonextensive statistics, and correspondingly the probability distribution function is found to be the power-law *q*-distribution. Up to now, nonextensive statistics has been developed as a very useful statistical theory for describing various complex systems including the long-range interacting astrophysical and space plasma systems with power-law distributions [2], among which plasma systems and self-gravitating systems become the most important application fields [3-9].

In the nonextensive kinetic theory [3,8,10], the velocity *q*-distribution function can be expressed as

$$f_q(\mathbf{v}) = nB_q \left(\frac{m}{2\pi k_B T}\right)^{\frac{3}{2}} \left[1 - (1-q)\frac{m(\mathbf{v}-\mathbf{u})^2}{2k_B T}\right]^{1/(1-q)}, \qquad (4)$$

where *n*, *T*, *m* and **u** is the density, the temperature, the particle mass and the bulk velocity of the system, respectively. $B_q$ is a *q*-dependent normalization constant given by

$$B_q = \begin{cases} (1-q)^{\frac{1}{2}}(3-q)(5-3q)\dfrac{\Gamma\left(\frac{1}{2}+\frac{1}{1-q}\right)}{4\Gamma\left(\frac{1}{1-q}\right)}, & 0 < q < 1, \\[2ex] (q-1)^{\frac{3}{2}}\dfrac{\Gamma\left(\frac{1}{q-1}\right)}{\Gamma\left(\frac{1}{q-1}-\frac{3}{2}\right)}, & q > 1. \end{cases}$$

When we take limit *q*→1, the velocity *q*-distribution (4) returns to the Maxwellian velocity distribution. In general, the temperature *T*, the density *n* and the bulk velocity **u** in a nonequilibrium complex plasma should be considered to be space inhomogeneous, i.e., *T*=*T*(*r*), *n*=*n*(*r*) and **u**=**u**(*r*), and they can vary with time.

As we known, plasma is a complex system with electromagnetic interactions, where the Coulomb force between charged particles is the long-range interaction. In some physical situations when the plasma has non-Maxwellian distributions or the power-law distributions, such as the *κ*-distributions and the *q*-distributions etc, the plasma physical properties are obviously different from those described under the framework of classical B-G statistics. The transport coefficients in the power-law *κ*-distributed plasma were first studied under a simplified Lorentz model [11]. Most recently, some transport properties in the weakly ionized plasma with power-law distribution were studied, such as the diffusion in the *κ*-distributed plasma [12], the viscosity in the *q*-distributed plasma [13], and the transport properties of the full ionized plasma with the *κ*-distribution [14,15]. In recent years, the power-law *κ*-distributions and the *κ*-like distributions in astrophysical and space plasmas have aroused widespread interest and concern [16]. And it has been found that these power-law distributed plasmas may be studied under the framework of nonextensive statistics if we take the simple transformation between the parameters *q* and *κ*, such as *κ*=(*q*-1)$^{-1}$ or *q*=1+*κ*$^{-1}$. More exactly, if the transformations are made [8] by both the parameter (*q*-1)$^{-1}$=*κ*+1 and the temperature *T* =(7-5*q*)$\tilde{T}$/2, the *κ*-distributions is equal to the *q*-distributions in nonextensive statistics, where $\tilde{T}$ is the temperature in the *κ*-distributions.

The purpose of this paper is to study the diffusion, thermal diffusion, thermal conductivity and thermoelectric coefficients in the weakly ionized plasma with *q*-distributions in nonextensive statistics. This paper is organized as follows. In Sec. 2, we introduce the transport equations for weakly ionized plasma with $q$-distributions in nonextensive statistics. In Sec. 3, we study expressions of the diffusion and thermal diffusion coefficients of charged particles in the q-distributed plasma. In Sec. 4, we study the thermal conductivity and thermoelectric coefficients of charged particles in the q-distributed plasma. Finally in Sec. 5, we give the conclusion.

**2. The transport equations for weakly ionized plasma with $q$-distributions**

Transport processes in a nonequilibrium thermodynamic system involve a variety of thermodynamic "fluxes", such as particle flow, heat flow and electric current etc. These "fluxes" are driven by the corresponding thermodynamic gradients; e.g. the particle flow is driven by density gradient, the heat flow is driven by temperature gradient, and the electric current is driven by electric potential gradient etc. According to the macroscopic thermodynamic laws for a multi-component fluid, the diffusion flow vector for the $\alpha$th component can be expressed [17] as

$$\mathbf{J}_{\alpha,D} = -D_\alpha \nabla n_\alpha - n_\alpha D_{T\alpha} \nabla T_\alpha, \tag{5}$$

where $D_{q,\alpha}$ is the diffusion coefficient, $D_{T\alpha}$ is the thermal diffusion coefficient, $n_\alpha$ is density and $T_\alpha$ is temperature. The ratio of the diffusion and thermal diffusion coefficients is called Soret coefficient $S_{T\alpha}$, i.e.,

$$S_{T\alpha} = \frac{D_{T\alpha}}{D_\alpha}. \tag{6}$$

Also, the heat flux vector for the $\alpha$th component can be expressed [17] as

$$\mathbf{J}_{\alpha,h} = -\lambda_\alpha \nabla T_\alpha - n_\alpha D_{T\alpha} \left(\frac{\partial \mu_\alpha}{\partial n_\alpha}\right)_{T,P} \frac{T_\alpha}{C_2} \nabla n_\alpha + \chi_\alpha T_\alpha \mathbf{J}_\alpha, \tag{7}$$

where $\lambda_\alpha$ is heat conductivity coefficient, $\mu_\alpha$ is chemical potential, $\chi_\alpha$ is thermoelectric coefficient, $\mathbf{J}_\alpha$ is the electric current density vector and $C_2$ is the mass fraction of the second component, e.g.,

$$C_2 = \frac{m_e n_e}{m_e n_e + m_i n_i} \text{ for electrons, and } C_2 = \frac{m_i n_i}{m_i n_i + m_e n_e} \text{ for ions.}$$

The weakly ionized plasma refers to the plasma whose ionization degree is less than 1%, where number of charged particles is small. So the deflections of charged particles are caused mainly by collision with the neutral atom rather than multiple scattering with other charged particles. The near collisions between the charged and neutral particles of the background gas are the main mechanism to determine the transport processes for charged particles in the plasma. If there are electrons, ions and neutral particles in weakly ionized plasma, because mass of neutral particles is relatively large as compared with electrons and ions, they are considered to be stationary static state. The Boltzmann form of the collision integral together with further simplifications can be used to study the transport properties of weakly ionized plasma. In nonextensive kinetics, the statistical description of this multi-body system can be given by the generalized Boltzmann equation [12,13],

$$\frac{\partial f_\alpha}{\partial t} + \mathbf{v} \cdot \frac{\partial f_\alpha}{\partial \mathbf{r}} + \frac{\mathbf{F}_\alpha}{m_\alpha} \cdot \frac{\partial f_\alpha}{\partial \mathbf{v}} = C_q(f_\alpha), \tag{8}$$

where $f_\alpha \equiv f_\alpha(\mathbf{r},\mathbf{v},t)$ is a single-particle velocity distribution function at time $t$, velocity $\mathbf{v}$, and position $\mathbf{r}$, the subscript $\alpha$ denotes the electron and the ion, $\alpha = e, i$, respectively, and $\mathbf{F}_\alpha$ is the external field force. Here we appoint that $\mathbf{r} = (r_x, r_y, r_z) \equiv (x, y, z)$. The term on the right-hand side $C_q$ is the nonextensive $q$-collision term, which represents the change in $f_\alpha$ due to the collisions. As a generalization, the $q$-collision term should (can) contain those collision terms given in the traditional statistics, for different physical situations. It is verified [18] that the velocity $q$-distribution (4) is a stationary solution of Eq.(8), and if the $q$-H theorem is satisfied, the $q$-collision term $C_q$ vanishes and $f_\alpha(\mathbf{r},\mathbf{v},t)$ will evolve irreversibly towards the velocity $q$-distribution.

In weakly ionized plasma, because the neutral particles are heavy as compared with electrons and ions, they can be regarded as static and homogeneous distribution and we mainly consider the collisions between charged particles and neutral particles and neglect the collision between the charged particles. For this reason, Krook collision model [19,12] can be generalized in nonextensive kinetics as

$$C_q(f_\alpha) = -\frac{1}{\tau_\alpha}\left(f_\alpha - f_{q,\alpha}^{(0)}\right) = -\nu_\alpha\left(f_\alpha - f_{q,\alpha}^{(0)}\right), \tag{9}$$

where $\tau_\alpha(v)$ is the mean time of collisions between the charged and neutral particle and here it can be assumed to be a constant. Or if $\nu_\alpha$ is the mean collision frequency, we have that $\nu_\alpha = (\tau_\alpha)^{-1}$. $f_{q,\alpha}^{(0)}$ is a $q$-equilibrium solution of Eq.(8).

Usually, in the first-order approximation of Chapman-Enskog expansion, we can write the solution of Eq.(8), the velocity distribution function, as the following form:

$$f_\alpha = f_{q,\alpha}^{(0)} + f_{q,\alpha}^{(1)}, \tag{10}$$

where $f_{q,\alpha}^{(1)}$ is a first-order small disturbance about the stationary $q$-distribution $f_{q,\alpha}^{(0)}$. In the case of smooth flow plasma without magnetic field, according to the power-law velocity $q$-distribution (4), we write that for the $\alpha$th component,

$$f_{q,\alpha}^{(0)}(\mathbf{r},\mathbf{v}) = n_\alpha B_{q,\alpha}\left(\frac{m_\alpha}{2\pi k_B T_\alpha}\right)^{3/2}\left[1 - (1-q_\alpha)\frac{m_\alpha}{2k_B T_\alpha}(\mathbf{v}-\mathbf{u})^2\right]^{1/(1-q_\alpha)}, \tag{11}$$

with

$$B_{q,\alpha} = \begin{cases} (1-q_\alpha)^{\frac{1}{2}}(3-q_\alpha)(5-3q_\alpha)\dfrac{\Gamma\left(\frac{1}{2}+\frac{1}{1-q_\alpha}\right)}{4\Gamma\left(\frac{1}{1-q_\alpha}\right)}, & \text{for } 0 < q_\alpha \leq 1. \\[2ex] (q_\alpha-1)^{\frac{3}{2}}\dfrac{\Gamma\left(\frac{1}{q_\alpha-1}\right)}{\Gamma\left(\frac{1}{q_\alpha-1}-\frac{3}{2}\right)}, & \text{for } q_\alpha \geq 1. \end{cases}$$

Substituting Eqs. (9)-(11) into Eq. (8), we have that

$$\left(\frac{\partial}{\partial t}+\mathbf{v}\cdot\frac{\partial}{\partial \mathbf{r}}+\frac{\mathbf{F}_\alpha}{m_\alpha}\cdot\frac{\partial}{\partial \mathbf{v}}\right)\left(f_{q,\alpha}^{(0)}+f_{q,\alpha}^{(1)}\right)=-\nu_\alpha f_{q,\alpha}^{(1)}. \tag{12}$$

Transport processes are all studied in a steady state so that $\partial f_\alpha/\partial t=0$. And the first-order small disturbance satisfies $f_{q,\alpha}^{(1)} \ll f_{q,\alpha}^{(0)}$, so we can neglect $f_{q,\alpha}^{(1)}$ on the left side of Eq.(12). Thus following the line in the textbooks [19], we can obtain the first-order approximation expression of the distribution function,

$$f_{q,\alpha}^{(1)}=-\frac{1}{\nu_\alpha}\left(\mathbf{v}\cdot\frac{\partial}{\partial \mathbf{r}}+\frac{\mathbf{F}_\alpha}{m_\alpha}\cdot\frac{\partial}{\partial \mathbf{v}}\right)f_{q,\alpha}^{(0)}. \tag{13}$$

Therefore from Eq.(10) we write the velocity distribution function,

$$f_\alpha = f_{q,\alpha}^{(0)} - \frac{1}{\nu_\alpha}\left(\mathbf{v}\cdot\frac{\partial}{\partial \mathbf{r}}+\frac{\mathbf{F}_\alpha}{m_\alpha}\cdot\frac{\partial}{\partial \mathbf{v}}\right)f_{q,\alpha}^{(0)}, \tag{14}$$

where the external field force is $\mathbf{F}_\alpha = Q_\alpha \mathbf{E}$ with the electric field intensity $\mathbf{E}$ and the electric charge $Q_\alpha$ for $\alpha$th component in the plasma, e.g., $Q_e=-e$ for the electron and $Q_i=Ze$ for the ion.

## 3. The diffusion and thermal diffusion coefficients

In the plasma, using (14) the diffusion flow density vector for the $\alpha$th component can be expressed by the velocity distribution function [19] as

$$\begin{aligned}\mathbf{J}_\alpha &= \int \mathbf{v} f_\alpha d\mathbf{v} \\ &= \int \mathbf{v} f_{q,\alpha}^{(0)} d\mathbf{v} - \int \mathbf{v}\, d\mathbf{v}\left(\frac{\mathbf{v}}{\nu_\alpha}\cdot\frac{\partial f_{q,\alpha}^{(0)}}{\partial \mathbf{r}}+\frac{\mathbf{F}_\alpha}{\nu_\alpha m_\alpha}\cdot\frac{\partial f_{q,\alpha}^{(0)}}{\partial \mathbf{v}_\alpha}\right).\end{aligned} \tag{15}$$

Because $f_{q,\alpha}^{(0)}$ is an even function about velocity $\mathbf{v}$, the first integral on the right-hand side of Eq. (15) is equal to zero. Then the diffusion flow density vector becomes

$$\mathbf{J}_\alpha = -\int \mathbf{v}\, d\mathbf{v}\left(\frac{\mathbf{v}}{\nu_\alpha}\cdot\frac{\partial f_{q,\alpha}^{(0)}}{\partial \mathbf{r}}+\frac{\mathbf{F}_\alpha}{\nu_\alpha m_\alpha}\cdot\frac{\partial f_{q,\alpha}^{(0)}}{\partial \mathbf{v}}\right), \tag{16}$$

where the first term is due to thermal motion denoted by $\mathbf{J}_{\alpha,D}$ and the second term is due to the electric field driven particle migration flow denoted by $\mathbf{\Gamma}_{\alpha,E}$.

In the nonequilibrium plasma, density and temperature are generally considered to be space inhomogeneous, i.e., $n_\alpha=n_\alpha(\mathbf{r})$ and $T_\alpha=T_\alpha(\mathbf{r})$. For the first term in Eq.(16), we therefore have that

$$\begin{aligned}\mathbf{J}_{\alpha,D} &= -\int \mathbf{v}\left(\frac{\mathbf{v}}{\nu_\alpha}\cdot\frac{\partial}{\partial \mathbf{r}}f_{q,\alpha}^{(0)}\right)d\mathbf{v} \\ &= -\int d\mathbf{v}\, \mathbf{v}\left\{\frac{1}{\nu_\alpha}\mathbf{v}\cdot\left[\frac{1}{n_\alpha}\nabla n_\alpha - \frac{3}{2}\frac{1}{T_\alpha}\nabla T_\alpha + \frac{m_\alpha v^2}{2k_B T_\alpha^2 - (1-q_\alpha)m_\alpha v^2 T_\alpha}\nabla T_\alpha\right]f_{q,\alpha}^{(0)}\right\} \\ &= -\frac{4\pi}{3\nu_\alpha n_\alpha}\nabla n_\alpha \int v^4 f_{q,\alpha}^{(0)} dv + \frac{2\pi}{\nu_\alpha T_\alpha}\nabla T_\alpha\left[\int v^4 f_{q,\alpha}^{(0)} dv - \frac{2m_\alpha}{3}\int \frac{v^6 f_{q,\alpha}^{(0)}\, dv}{2k_B T_\alpha - (1-q_\alpha)m_\alpha v^2}\right].\end{aligned}$$

(17)

For $q_\alpha > 1$, substituting the distribution function (11) into Eq.(17), after calculation of the integrals (see Appendix A) we obtain that

$$\mathbf{J}_{\alpha,D} = -\frac{4\pi}{3\nu_\alpha n_\alpha}\nabla n_\alpha \int_0^\infty v^4 f_{q,\alpha}^{(0)} dv$$

$$+\frac{2\pi}{\nu_\alpha T_\alpha}\nabla T_\alpha \left[\int_0^\infty v^4 f_{q,\alpha}^{(0)} dv - \frac{2m_\alpha}{3}\int_0^\infty \frac{v^6 f_{q,\alpha}^{(0)}}{2k_B T_\alpha - (1-q_\alpha)m_\alpha v^2} dv\right]$$

$$= -\frac{2k_B T_\alpha}{\nu_\alpha m_\alpha (7-5q_\alpha)}\nabla n_\alpha - \frac{2n_\alpha k_B}{\nu_\alpha m_\alpha (7-5q_\alpha)}\nabla T_\alpha, \quad 1 < q_\alpha < \frac{7}{5}. \tag{18}$$

For $0 < q_\alpha < 1$, because there is a thermal cutoff on the maximum value allowed for velocity, $v_{max} = \sqrt{2k_B T_\alpha / m_\alpha (1-q_\alpha)}$, in the distribution functions, we have that

$$\mathbf{J}_{\alpha,D} = -\frac{4\pi}{3\nu_\alpha n_\alpha}\nabla n_\alpha \int_0^{v_{max}} v^4 f_{q,\alpha}^{(0)} dv + \frac{2\pi}{\nu_\alpha T_\alpha}\nabla T_\alpha \left[\int_0^{v_{max}} v^4 f_{q,\alpha}^{(0)} dv\right.$$

$$\left.-\frac{2m_\alpha}{3}\int_0^{v_{max}} \frac{v^6 f_{q,\alpha}^{(0)}}{2k_B T_\alpha - (1-q_\alpha)m_\alpha v^2} dv\right]. \tag{19}$$

After calculation of the integrals (see Appendix A) we obtain that

$$\mathbf{J}_{\alpha,D} = -\frac{2k_B T_\alpha}{\nu_\alpha m_\alpha (7-5q_\alpha)}\nabla n_\alpha - \frac{2n_\alpha k_B}{\nu_\alpha m_\alpha (7-5q_\alpha)}\nabla T_\alpha, \quad 0 < q_\alpha < 1. \tag{20}$$

Comparing Eqs.(18) and (20) with Eq.(5), we find that the diffusion coefficient is

$$D_{q,\alpha} = \frac{2k_B T_\alpha}{\nu_\alpha m_\alpha (7-5q_\alpha)}, \quad 0 < q_\alpha < \frac{7}{5}, \tag{21}$$

and the thermal diffusion coefficient is

$$D_{q,T\alpha} = \frac{2k_B}{\nu_\alpha m_\alpha (7-5q_\alpha)}, \quad 0 < q_\alpha < \frac{7}{5}. \tag{22}$$

It is showed that in the weakly ionized plasma with the power-law $q$-distributions, the diffusion and thermal diffusion coefficients (21) and (22) are significantly depended on the nonextensive parameters, and when we take the limit $q_\alpha \to 1$, (21) and (22) recover the standard forms in a Maxwellian distribution.

For the second term in Eq.(16), i.e. the electric field driven particle migration flow, we have that

$$\mathbf{\Gamma}_{\alpha,E} = -\int \mathbf{v}\, d\mathbf{v} \left(\frac{\mathbf{F}_\alpha}{\nu_\alpha m_\alpha} \cdot \frac{\partial f_{q,\alpha}^{(0)}}{\partial \mathbf{v}}\right)$$

$$= \frac{4\pi Q_\alpha \mathbf{E}}{\nu_\alpha m_\alpha}\int v^2 f_{q,\alpha}^{(0)} dv = \frac{n_\alpha Q_\alpha \mathbf{E}}{\nu_\alpha m_\alpha}. \tag{23}$$

If the mobility for the $\alpha$th component in the plasma is denoted as $\mu_\alpha$, according to the macroscopic law [19],
$$\mathbf{\Gamma}_{\alpha,E} = \mu_{q,\alpha} n_\alpha \mathbf{E}, \tag{24}$$
we find the mobility
$$\mu_{q,\alpha} = \frac{Z_\alpha e}{m_\alpha v_\alpha}, \tag{25}$$
where $Q_\alpha = Z_\alpha e$ is used. It is clear that the mobility does not depend on the $q$-parameters, so in the weakly ionized plasma with the power-law $q$-distributions, the mobility is the same as the standard form of the plasma with a Maxwell distribution,.

From (20) and (21), we can write that
$$\frac{D_{q,\alpha}}{D_{1,\alpha}} = \frac{D_{q,T\alpha}}{D_{1,T\alpha}} = \frac{2}{7-5q_\alpha}, \quad 0 < q_\alpha < \frac{7}{5}, \tag{26}$$
where $D_{1,\alpha}$ and $D_{1,T\alpha}$ are respectively the diffusion and thermal diffusion coefficient of $\alpha$th component in the case of plasma with a Maxwellian distribution. Therefore, in fig.1, we demonstrated a numerical analysis on dependence of the diffusion and thermal diffusion coefficient on the $q$-parameter, relative to that in the case of the Maxwellian distribution.

The figure showed clearly the significant effect of the nonextensivity on the diffusion and thermal diffusion coefficient of electrons and ions in the plasma. They increase monotonously as the $q$-parameter increase, and if the parameter is $q < 1$ they are less than those in the case of the plasma with the Maxwellian distribution, if $q > 1$ they are more than those in the case of the plasma with the Maxwellian distribution.

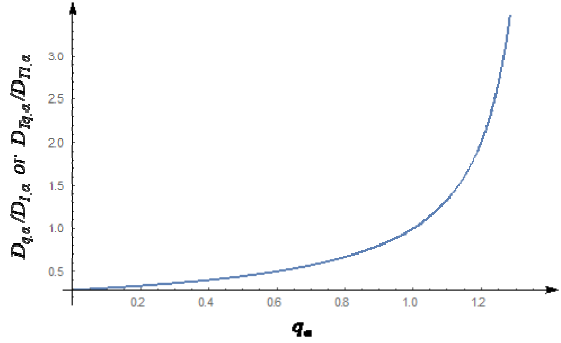

Fig.1. Dependence of the diffusion and thermal diffusion
coefficient on the $q$-parameter

## 4. The thermal conductivity and thermoelectric coefficients

The heat flux vector for the $\alpha$th component in the plasma is defined as
$$\mathbf{J}_{\alpha,h} = \int \frac{1}{2} m_\alpha v^2 \mathbf{v} f_\alpha \, d\mathbf{v}.$$

And therefore we have
$$\mathbf{J}_{\alpha,h} = \frac{1}{2} m_\alpha \int v^2 \mathbf{v} \left( f_{q,\alpha}^{(0)} + f_{q,\alpha}^{(1)} \right) d\mathbf{v} = \frac{1}{2} m_\alpha \int v^2 \mathbf{v} f_{q,\alpha}^{(1)} \, d\mathbf{v}, \tag{27}$$
where $\int v^2 \mathbf{v} f_{q,\alpha}^{(0)} \, d\mathbf{v} = 0$ has been used because the integrand is an even function of velocity. Using (13) we have that

$$\mathbf{J}_{\alpha,h} = -\frac{m_\alpha}{2\nu_\alpha}\int v^2 \mathbf{v}\left(\mathbf{v}\cdot\frac{\partial f_{q,\alpha}^{(0)}}{\partial \mathbf{r}} + \frac{\mathbf{F}_\alpha}{m_\alpha}\cdot\frac{\partial f_{q,\alpha}^{(0)}}{\partial \mathbf{v}}\right)d\mathbf{v}, \quad (28)$$

Further, calculating the first integral term in (28) it becomes

$$\mathbf{J}_{\alpha,h} = -\frac{m_\alpha}{2\nu_\alpha}\int v^2 \mathbf{v}\, d\mathbf{v}\left[\mathbf{v}\cdot\frac{\nabla n_\alpha}{n_\alpha}f_{q,\alpha}^{(0)} + \mathbf{v}\cdot\frac{\nabla T_\alpha}{T_\alpha}\left(-\frac{3}{2} + \frac{m_\alpha v^2}{2k_B T_\alpha - (1-q_\alpha)m_\alpha v^2}\right)f_{q,\alpha}^{(0)} + \frac{\mathbf{F}_\alpha}{m_\alpha}\cdot\frac{\partial f_{q,\alpha}^{(0)}}{\partial \mathbf{v}}\right]. \quad (29)$$

For $q_\alpha > 1$, the integrals should be written as

$$\mathbf{J}_{\alpha,h} = -\frac{m_\alpha}{2n_\alpha\nu_\alpha}\int_0^\infty v^2(\mathbf{vv}\cdot\nabla n_\alpha)f_{q,\alpha}^{(0)}d\mathbf{v} + \frac{3m_\alpha}{4\nu_\alpha T_\alpha}\int_0^\infty v^2(\mathbf{vv}\cdot\nabla T_\alpha)f_{q,\alpha}^{(0)}d\mathbf{v}$$

$$-\frac{m_\alpha^2}{2\nu_\alpha}\int_0^\infty \frac{v^4(\mathbf{vv}\cdot\nabla T_\alpha)f_{q,\alpha}^{(0)}}{2k_B T_\alpha^2 - (1-q_\alpha)m_\alpha v^2 T_\alpha}d\mathbf{v} - \frac{Q_\alpha\mathbf{E}}{2\nu_\alpha}\int_0^\infty v^2\mathbf{v}\cdot\frac{\partial f_{q,\alpha}^{(0)}}{\partial \mathbf{v}}d\mathbf{v}. \quad (30)$$

After the integrals in (29) are calculated (see Appendix B), we obtain

$$\mathbf{J}_{\alpha,h} = -\frac{5k_B T_\alpha}{\nu_\alpha m_\alpha(5q_\alpha - 7)}\left[\frac{2k_B T_\alpha}{(7q_\alpha - 9)}\nabla n_\alpha + \frac{4n_\alpha k_B}{(7q_\alpha - 9)}\nabla T_\alpha + n_\alpha Q_\alpha\mathbf{E}\right], \quad 1 < q_\alpha < \frac{9}{7}. \quad (31)$$

For $0 < q_\alpha < 1$, because there is a thermal cutoff on the maximum value allowed for velocity, $v_{\max} = \sqrt{2k_B T_\alpha/m_\alpha(1-q_\alpha)}$, in the distribution functions, after the integrals in (29) are calculated (see Appendix B) we have that

$$\mathbf{J}_{\alpha,h} = -\frac{m_\alpha}{2n_\alpha\nu_\alpha}\int_0^{v_{\max}} v^2(\mathbf{vv}\cdot\nabla n_\alpha)f_{q,\alpha}^{(0)}d\mathbf{v} + \frac{3m_\alpha}{4\nu_\alpha T_\alpha}\int_0^{v_{\max}} v^2(\mathbf{vv}\cdot\nabla T_\alpha)f_{q,\alpha}^{(0)}d\mathbf{v}$$

$$-\frac{m_\alpha^2}{2\nu_\alpha}\int_0^{v_{\max}} \frac{v^4(\mathbf{vv}\cdot\nabla T_\alpha)f_{q,\alpha}^{(0)}}{2k_B T_\alpha^2 - (1-q_\alpha)m_\alpha v^2 T_\alpha}d\mathbf{v} - \frac{Q_\alpha\mathbf{E}}{2\upsilon_\alpha}\int_0^{v_{\max}} v^2\mathbf{v}\cdot\frac{\partial f_{q,\alpha}^{(0)}}{\partial \mathbf{v}}d\mathbf{v}$$

$$= -\frac{5k_B T_\alpha}{\nu_\alpha m_\alpha(5q_\alpha - 7)}\left[\frac{2k_B T_\alpha}{(7q_\alpha - 9)}\nabla n_\alpha + \frac{4n_\alpha k_B}{(7q_\alpha - 9)}\nabla T_\alpha + n_\alpha Q_\alpha\mathbf{E}\right]. \quad (32)$$

Comparing Eqs.(31) and (32) with the macroscopic law Eq.(7), we find the expressions of thermal conductivity coefficient,

$$\lambda_{q,\alpha} = \frac{20n_\alpha k_B^2 T_\alpha}{\nu_\alpha m_\alpha(5q_\alpha - 7)(7q_\alpha - 9)}, \quad 0 < q_\alpha < \frac{9}{7}, \quad (33)$$

and thermoelectric coefficient,

$$\chi_{q,\alpha} = \frac{5k_B}{Q_\alpha(5q_\alpha - 7)}, \quad 0 < q_\alpha < \frac{9}{7}, \quad (34)$$

where the electric current density vector $\mathbf{J}_{\alpha,e} = Q_\alpha\mathbf{\Gamma}_{\alpha,E}$ with the electric field driven particle migration flow in (23) has been used. It is shown that in the weakly ionized plasma with the $q$-distributions, thermal conductivity coefficient and thermoelectric coefficient (33) and (34) depend significantly on the nonextensive $q$-parameters, and when we take the limit $q_\alpha \to 1$, (33) and (34) recover the standard forms in a Maxwellian distribution.

From (33) and (34), we can write that

$$\frac{\lambda_{q,\alpha}}{\lambda_{1,\alpha}} = \frac{4}{(5q_\alpha - 7)(7q_\alpha - 9)} \quad \text{and} \quad \frac{\chi_{q,\alpha}}{\chi_{1,\alpha}} = \frac{2}{7 - 5q_\alpha}, \quad 0 < q_\alpha < \frac{9}{7}, \tag{35}$$

where $\lambda_{1,\alpha}$ and $\chi_{1,\alpha}$ are respectively the thermal conductivity and thermoelectric coefficient of $\alpha$th component in the case of the plasma with a Maxwellian distribution. Therefore, in fig.2 (a) and (b), respectively, we demonstrated a numerical analysis on dependence of these two transport coefficients on the $q$-parameter, relative to that in the case of the plasma with a Maxwellian distribution.

The figures (a) and (b) showed clearly the significant effect of the nonextensivity on the thermal conductivity and thermoelectric coefficient of electrons and ions in the plasma. They increase monotonously as the $q$-parameter increase, and if the parameter is $q < 1$ they are less than those in the case of the plasma with the Maxwellian distribution, if $q > 1$ they are more than those in the case of the plasma with the Maxwellian distribution.

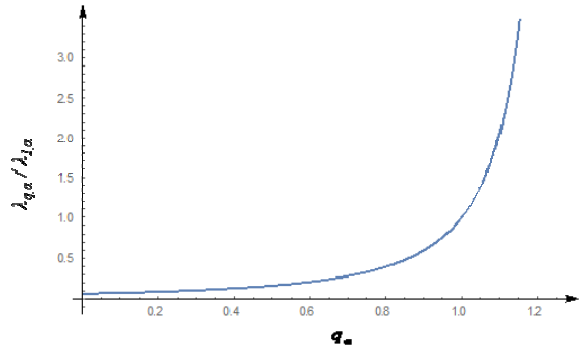

(a)

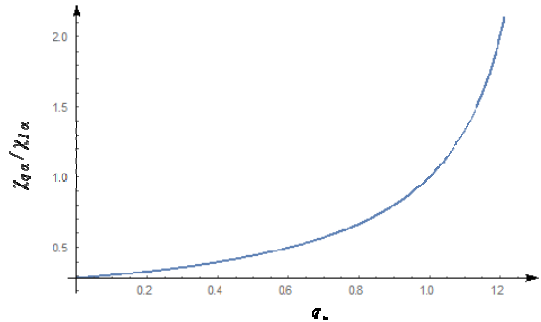

(b)

Fig.2. Dependence of the thermal conductivity coefficient in (a) and the thermoelectric coefficient in (b) on the $q$-parameter.

## 5. Conclusion and discussion

In conclusion, we have studied the transport properties of charged particles in the weakly ionized plasma with the power-law $q$-distributions and without the magnetic field, including the diffusion, thermal diffusion, thermal conductivity and thermoelectric coefficients for the electrons and ions. Electrons and ions in the plasma have different $q$-parameters, particle density and temperature. We introduced the transport equations for the weakly ionized plasma with the power-law $q$-distributions in nonextensive statistics. Under this framework, we derive expressions of the diffusion coefficient, thermal diffusion coefficient, thermal conductivity coefficient and thermoelectric coefficient of electrons and ions, which are given by (21), (22), (33) and (34), respectively. We show that these transport coefficients depend significantly on the nonextensive $q$-parameters in the plasma with the power-law $q$-distributions, and when we take the limit $q_\alpha \to 1$, they can recover the standard forms in a Maxwell distribution.

Numerical analyses find that the four transport coefficients all increase monotonously as the $q$-parameters increase, and if the parameter is $q < 1$ they are all less than those in the case of the plasma with the Maxwellian distribution, if $q > 1$ they are all more than those in the case of the plasma with the Maxwellian distribution.

On the nonextensivity, i.e., the degree of the $q$-parameter deviation from unity $q \neq 1$, in the nonequilibrium plasma with power-law $q$-distributions, its physical meaning can be understood by the relation [3],

$$k_B \nabla T_\alpha = (q_\alpha - 1) Q_\alpha \nabla \varphi_c, \qquad (36)$$

where $\varphi_c$ is the Coulomb potential. Thus, the nonextensive $q$-parameter different from unity represented correctly the nonequilibrium properties of the plasma with the Coulomb long-range interactions between the charged particles. These transport coefficients obtained in this paper can be applicable to study the weakly ionized plasmas with the above physical properties.

Bezerra et al discussed the transport of a general gas with a velocity $q$-distribution [20]. They assumed the particle number density to be constant and derived a similar expression of thermal conductivity coefficient only. In the present paper, based on the nonuniform particle density in the weakly ionized plasma with the power-law $q$-distributions, we derived the above four transport coefficients for the electrons and ions respectively in the plasma, which will inspire us the way to study further the transport properties of the plasmas with the power-law $q$-distributions and in more different and complex physical situations.

## Acknowledgments

This work is supported by the National Natural Science Foundation of China under Grant No. 11775156.

## Appendix A

For $q_\alpha > 1$, Eq.(17) is written as

$$\mathbf{J}_{\alpha,D} = \left(-\frac{4\pi}{3\nu_\alpha n_\alpha}\nabla n_\alpha + \frac{2\pi}{\nu_\alpha T_\alpha}\nabla T_\alpha\right)\int_0^\infty v^4 f_{q,\alpha}^{(0)} dv - \frac{4\pi m_\alpha}{3\nu_\alpha T_\alpha}\nabla T_\alpha \int_0^\infty \frac{v^6 f_{q,\alpha}^{(0)}}{2k_B T_\alpha - (1-q_\alpha)m_\alpha v^2} dv$$

$$= -(T_\alpha \nabla n_\alpha + n_\alpha \nabla T_\alpha)\frac{k_B B_{q,\alpha}}{\nu_\alpha m_\alpha (q_\alpha - 1)^{5/2}} \frac{\Gamma\left(-\frac{5}{2} + \frac{1}{q_\alpha - 1}\right)}{\Gamma\left(\frac{1}{q_\alpha - 1}\right)}$$

$$= -\frac{2k_B T_\alpha}{\nu_\alpha m_\alpha (7-5q_\alpha)}\nabla n_\alpha - \frac{2n_\alpha k_B}{\nu_\alpha m_\alpha (7-5q_\alpha)}\nabla T_\alpha, \quad 0 < q_\alpha < \frac{7}{5}.$$

For $0 < q_\alpha < 1$, if we use $a = \sqrt{\frac{2k_B T_\alpha}{m_\alpha (1-q_\alpha)}}$, Eq. (19) is written as

$$\mathbf{J}_{\alpha,D} = -\frac{4\pi}{3\nu_\alpha n_\alpha}\nabla n_\alpha \int_0^a v^4 f_{q,\alpha}^{(0)} dv + \frac{2\pi}{\nu_\alpha T_\alpha}\nabla T_\alpha \int_0^a v^4 f_{q,\alpha}^{(0)} dv$$

$$-\frac{4\pi m_\alpha}{3\nu_\alpha T_\alpha}\nabla T_\alpha \int_0^a \frac{v^6 f_{q,\alpha}^{(0)}}{2k_B T_\alpha - (1-q_\alpha)m_\alpha v^2} dv$$

$$= \left(-T_\alpha \nabla n_\alpha + \frac{3n_\alpha}{2}\nabla T_\alpha\right)\frac{k_B B_{q,\alpha}}{\nu_\alpha m_\alpha (1-q_\alpha)^{5/2}} \frac{\Gamma\left(1+\frac{1}{1-q_\alpha}\right)}{\Gamma\left(\frac{7}{2}+\frac{1}{1-q_\alpha}\right)} - \frac{5n_\alpha k_B B_{q,\alpha}}{2\nu_\alpha m_\alpha (1-q_\alpha)^{7/2}} \frac{\Gamma\left(\frac{1}{1-q_\alpha}\right)}{\Gamma\left(\frac{7}{2}+\frac{1}{1-q_\alpha}\right)}\nabla T_\alpha$$

$$= -\frac{2k_B T_\alpha}{\nu_\alpha m_\alpha (7-5q_\alpha)}\nabla n_\alpha - \frac{2n_\alpha k_B}{\nu_\alpha m_\alpha (7-5q_\alpha)}\nabla T_\alpha.$$

**Appendix B**

If $q_\alpha > 1$, (30) is written as

$$\mathbf{J}_{\alpha,h} = -\frac{m_\alpha}{2n_\alpha \nu_\alpha}\int_0^\infty v^2 (\mathbf{vv}\cdot\nabla n_\alpha) f_{q,\alpha}^{(0)} d\mathbf{v} + \frac{3m_\alpha}{4\nu_\alpha T_\alpha}\int_0^\infty v^2 (\mathbf{vv}\cdot\nabla T_\alpha) f_{q,\alpha}^{(0)} d\mathbf{v}$$

$$-\frac{m_\alpha^2}{2\nu_\alpha}\int_0^\infty \frac{v^4 (\mathbf{vv}\cdot\nabla T_\alpha) f_{q,\alpha}^{(0)}}{2k_B T_\alpha^2 - (1-q_\alpha)m_\alpha v^2 T_\alpha} d\mathbf{v} - \frac{Q_\alpha \mathbf{E}}{2\nu_\alpha}\int_0^\infty v^2 \mathbf{v}\cdot\frac{\partial}{\partial \mathbf{v}} f_{q,\alpha}^{(0)} d\mathbf{v}$$

$$= -\frac{5k_B^2 T_\alpha^2 B_{q,\alpha}}{\nu_\alpha m_\alpha (q_\alpha - 1)^{7/2}} \frac{\Gamma\left(-\frac{7}{2}+\frac{1}{q_\alpha - 1}\right)}{\Gamma\left(\frac{1}{q_\alpha - 1}\right)}\left(\frac{1}{2}\nabla n_\alpha + n_\alpha \nabla T_\alpha\right) + \frac{5n_\alpha k_B T_\alpha B_{q,\alpha} Q_\alpha}{2\nu_\alpha m_\alpha (q_\alpha - 1)^{5/2}} \frac{\Gamma\left(-\frac{5}{2}+\frac{1}{q_\alpha - 1}\right)}{\Gamma\left(\frac{1}{q_\alpha - 1}\right)}\mathbf{E}$$

$$= -\frac{5k_B T_\alpha}{\nu_\alpha m_\alpha (5q_\alpha - 7)}\left[\frac{2k_B T_\alpha}{(7q_\alpha - 9)}\nabla n_\alpha + \frac{4n_\alpha k_B}{(7q_\alpha - 9)}\nabla T_\alpha + n_\alpha Q_\alpha \mathbf{E}\right], \quad 1 < q_\alpha < \frac{9}{7}.$$

This is Eq.(31).

For $0 < q_\alpha < 1$, if we use $a = \sqrt{\frac{2k_B T_\alpha}{m_\alpha (1-q_\alpha)}}$, Eq.(29) is written as

$$\mathbf{J}_{\alpha,h} = -\frac{m_\alpha}{2n_\alpha \nu_\alpha}\int_0^a v^2 (\mathbf{vv}\cdot\nabla n_\alpha) f_{q,\alpha}^{(0)} d\mathbf{v} + \frac{3m_\alpha}{4\nu_\alpha T_\alpha}\int_0^a v^2 (\mathbf{vv}\cdot\nabla T_\alpha) f_{q,\alpha}^{(0)} d\mathbf{v}$$

$$-\frac{m_\alpha^2}{2\nu_\alpha}\int_0^a \frac{v^4(\mathbf{vv}\cdot\nabla T_\alpha)f_{q,\alpha}^{(0)}}{2k_B T_\alpha^2-(1-q_\alpha)m_\alpha v^2 T_\alpha}d\mathbf{v}+\frac{Q_\alpha \mathbf{E}}{2\upsilon_\alpha}\int_0^a v^2\mathbf{v}\cdot\frac{\partial f_{q,\alpha}^{(0)}}{\partial \mathbf{v}}d\mathbf{v}$$

$$=-\frac{5k_B^2 T_\alpha^2 B_{q,\alpha}}{\nu_\alpha m_\alpha(1-q_\alpha)^{9/2}}\frac{\Gamma\left(\dfrac{1}{1-q_\alpha}\right)}{\Gamma\left(\dfrac{9}{2}+\dfrac{1}{1-q_\alpha}\right)}\left(\frac{1}{2}\nabla n_\alpha+n_\alpha \nabla T_\alpha\right)-\frac{5n_\alpha Q_\alpha k_B T_\alpha B_{q,\alpha}}{2\nu_\alpha m_\alpha(1-q_\alpha)^{5/2}}\frac{\Gamma\left(1+\dfrac{1}{1-q_\alpha}\right)}{\Gamma\left(\dfrac{7}{2}+\dfrac{1}{1-q_\alpha}\right)}\mathbf{E}$$

$$=-\frac{5k_B T_\alpha}{\nu_\alpha m_\alpha(5q_\alpha-7)}\left[\frac{2k_B T_\alpha}{(7q_\alpha-9)}\nabla n_\alpha+\frac{4n_\alpha k_B}{(7q_\alpha-9)}\nabla T_\alpha+n_\alpha Q_\alpha \mathbf{E}\right],\ 0<q_\alpha<1.$$

This is Eq.(32).